\documentclass[12pt]{article}

\usepackage[margin=1in]{geometry}
\usepackage{graphicx}
\usepackage{booktabs}
\usepackage{natbib}
\usepackage{hyperref}
\usepackage{xcolor}
\usepackage{caption}
\usepackage{authblk}
\usepackage{amsmath}

\hypersetup{
    colorlinks=true,
    linkcolor=blue,
    citecolor=blue,
    urlcolor=blue,
    breaklinks=true
}
\title{Views on AI Existential Risk Before and After a Public Event at Harvard University}

\author[1]{Greg Kestin}
\author[2]{Nate Soares}
\affil[1]{Department of Physics, Harvard University, Cambridge, MA}
\affil[2]{Machine Intelligence Research Institute, Berkeley, CA}

\date{March 2026}

\begin{document}

\maketitle

\begin{abstract}
We report the results of identical pre- and post-event surveys given to attendees of a talk, two-sided conversation, and Q\&A centered around the book \textit{If Anyone Builds It, Everyone Dies} at Harvard University in March 2026, covering perceived probability of AI-caused extinction or severe disempowerment resulting from unimpeded AI development, confidence in those estimates, and global priority. Among the 89 matched participants, the post-event median estimate of the probability of existential risk from advanced AI was 70\%, and 96\% agreed that mitigating AI existential risk should be a global priority. Although these self-selected respondents' pre-event views were already high (50\% and 93\%, respectively) relative to results of similar surveys that were previously administered to experts and the general public, the event produced increases on all measures when considering the respondents in aggregate. The magnitudes of increases in risk probability were negatively correlated with prior familiarity with the topic: among attendees with little prior familiarity, 60\% shifted upward and none shifted downward, whereas among self-described experts, no respondents shifted upward and 20\% shifted downward. Self-reported confidence also increased significantly, and confidence shifts were positively correlated with probability shifts. These findings indicate that a structured public engagement event can meaningfully shift risk perceptions, particularly among newcomers to the topic.
\end{abstract}

\section{Introduction}

Public understanding of existential risk from advanced artificial intelligence remains an emerging area of study. While surveys of AI researchers have documented expert views on the probability of catastrophic AI outcomes \citep{grace2024}, and polls of the general public and university students have gauged general sentiment \citep{aipi2023, hirabayashi2024}, comparatively little work has examined how audiences with varying levels of prior familiarity assess AI existential risk and update their beliefs when exposed to structured arguments and discussion about these risks.

As AI capabilities advance rapidly---with AI systems demonstrating capabilities sufficient to win the International Mathematical Olympiad \citep{hubert2025, huang2025}, over a decade ahead of forecaster predictions \citep{metaculus2021} and producing novel contributions to theoretical physics \citep{guevara2026}---and as AI systems increasingly operate autonomously as agents in the world, understanding how the public perceives these risks becomes increasingly important.

On March 11, 2026, we conducted a pre/post survey at a \href{https://science.fas.harvard.edu/book-talks}{Harvard Science Book Talk} event (co-hosted by the Harvard Division of Science, Harvard Book Store, and Harvard Library). The event featured a talk by Nate Soares, president of the Machine Intelligence Research Institute (MIRI), based on the book \textit{If Anyone Builds It, Everyone Dies} \citep{soares2025}, followed by a structured conversation facilitated by Greg Kestin\footnote{Kestin teaches in Harvard's Physics Department and serves as the Associate Director of Science Education; his role and research have a focus on AI and its implications.} that addressed arguments about the danger of artificial superintelligence along with questions and discussion of those arguments.

We report the results of this survey, examining perceived risk, confidence, and prioritization of mitigating AI existential risk, how these shifted following the event, and how shifts varied by prior familiarity with the topic. We also compare our results to existing benchmarks: the 2023 Expert Survey on Progress in AI \citep{grace2024} for risk probability, and polls of the general public and university students \citep{aipi2023, hirabayashi2024} for global priority.

\section{Methods}

\subsection{Event format}

The event consisted of an approximately 13-minute talk by Soares summarizing the case that sufficiently advanced AI poses an extinction threat, followed by a moderated conversation of approximately 35 minutes. The conversation, facilitated by Kestin, was designed to be two-sided: while drawing out the speaker's arguments, Kestin also raised counterarguments, common questions, and alternative perspectives gathered beforehand from colleagues in and outside academia as well as from the AI safety community. For example, Kestin raised the question of whether large language models could carry inherent alignment advantages over earlier paradigms, since, for example, LLMs are trained on human data and \emph{may} therefore be better positioned to interpret human goals and values. The event concluded with an approximately 13-minute audience question-and-answer session. The event was recorded and is available at \url{https://youtu.be/TD9AFmJy_GY}.

\subsection{Survey design}

Identical surveys were administered before and after the event using anonymous polling. Participants accessed the survey via QR code or URL that was presented to them.

The pre-event survey contained four questions:

\begin{enumerate}
    \item \textbf{Probability of AI existential risk.} ``What probability do you put on future AI advances causing human extinction or similarly permanent and severe disempowerment of the human species? (Assume that companies developing AI technology proceed largely unimpeded by external intervention.)'' The probabilities were separated into seven bins that served as answer choices: Under 5\% (very unlikely or impossible), 5--20\% (unlikely), 20--40\% (possible but unlikely), 40--60\% (roughly a coin flip), 60--80\% (more likely than not), 80--95\% (very likely), and Above 95\% (near certain). This question was adapted from the 2023 Expert Survey on Progress in AI \citep{grace2024}, which asked AI researchers an analogous question with free-form numerical responses. We added a parenthetical clause asking respondents to assume that AI development proceeds largely unimpeded, and we used binned responses for ease of completion by a general audience.

    \item \textbf{Confidence.} ``How confident are you in your response to the previous question?'' The answer choices were a five-point scale from ``Not at all confident (pure guess)'' to ``Completely confident (firm, unchangeable view).''

    \item \textbf{Global priority.} ``Mitigating the risk of extinction from AI should be a global priority alongside other societal-scale risks such as pandemics and nuclear war.'' The answer choices were a five-point Likert scale from ``Strongly disagree'' to ``Strongly agree.'' This wording is taken directly from the Center for AI Safety (CAIS) statement on AI risk \citep{cais2023}.

    \item \textbf{Role/affiliation.} ``Which best describes your current role or affiliation?'' Options included undergraduate student, graduate/professional/postdoctoral student, faculty or academic researcher, university staff, work outside academia in a scientific field, work outside academia in a non-science field, retired, and other.
\end{enumerate}

The post-event survey repeated all four questions and added a fifth:

\begin{enumerate}
    \setcounter{enumi}{4}
    \item \textbf{Prior exposure} (post-event only). ``Before today, how much had you heard or read about arguments that advanced AI (e.g., superhuman AI) could pose an existential risk to humanity?'' The five answer choices were: Nothing at all, A little, A moderate amount, A great deal, Expert level.
\end{enumerate}

\subsection{Participants}

The event was an in-person public lecture at 6 PM, publicized primarily to the Harvard Faculty of Arts and Sciences community and subscribers to the Harvard Book Store mailing list. Attendance was voluntary. Of the approximately 180 attendees, 129 completed the pre-event survey and 113 completed the post-event survey. We matched 89 participants who completed both surveys. The unmatched participants---40 of whom only took the pre-event survey and 24 of whom only took the post-event survey---were excluded from the primary matched analysis.

The demographic and exposure composition of the matched sample ($n = 89$) is shown in Table~\ref{tab:demographics}.

\begin{table}[h]
\centering
\caption{Composition of the matched sample ($n = 89$) by role and prior exposure.}
\label{tab:demographics}
\begin{tabular}{lcc}
\toprule
\textbf{Role/affiliation} & $n$ & \% \\
\midrule
Work outside academia, scientific field & 23 & 26\% \\
Work outside academia, non-science field & 22 & 25\% \\
Retired & 13 & 15\% \\
Graduate/professional/postdoctoral student & 9 & 10\% \\
Faculty or academic researcher & 8 & 9\% \\
Other & 6 & 7\% \\
University staff & 4 & 4\% \\
Undergraduate student & 4 & 4\% \\
\midrule
\textbf{Prior exposure to AI risk topic} & $n$ & \% \\
\midrule
A great deal & 32 & 36\% \\
A moderate amount & 31 & 35\% \\
A little & 15 & 17\% \\
Expert level & 10 & 11\% \\
Nothing at all & 1 & 1\% \\
\bottomrule
\end{tabular}
\end{table}

\subsection{Analysis}

For the primary analysis, we used the 89 matched participants. Each participant's pre- and post-event responses to Q1 (probability) were compared to determine shift direction: whether the post-event response was higher, lower, or unchanged relative to the pre-event response. For numerical analyses, responses were coded to bin midpoints (2.5\%, 12.5\%, 30\%, 50\%, 70\%, 87.5\%, 97.5\%).

To test for overall shifts, we used Wilcoxon signed-rank tests. To assess the relationship between prior exposure and the magnitude of shifts, we used Spearman rank correlations. We report descriptive statistics stratified by prior exposure level (Q5).

In addition to $p$-values, we report likelihood ratios (LRs) comparing the best-fit hypothesis to the null. For shift analyses, the LR compares a model where the probability of upward shift equals the observed proportion to the null model where upward and downward shifts are equally likely. For correlations, the LR compares the observed correlation coefficient to zero. A likelihood ratio of $k{:}1$ indicates the data are $k$ times more likely under the best-fit hypothesis than under the null.

\subsection{Ethics}

This study involved voluntary, anonymous polling at a public lecture event. No personally identifiable information was collected. Participants were informed that their responses were anonymous and would be used to compare pre- and post-event views. The survey was administered as part of the event programming.\footnote{Given the anonymous, voluntary nature of the survey and the absence of personally identifiable information, this study is not deemed human subjects research.}

\section{Results}

\subsection{Overall views before and after the event}

Among the 89 matched participants, pre-event estimates of AI existential risk were high: the median response was in the 40--60\% bin (midpoint 50\%), the mean was 50.5\%, and 57\% of respondents placed risk at 40\% or higher. After the event, these estimates shifted upward: the median rose to the 60--80\% bin (midpoint 70\%), the mean increased to 56.9\%, and 65\% placed risk at 40\% or higher.

Pre-event confidence in risk estimates was moderate, with a mean of 2.72 on the 5-point scale (where 1 = ``not at all confident'' and 5 = ``completely confident''). After the event, mean confidence rose to 3.02, with the proportion reporting ``Very'' or ``Completely'' confident increasing from 17\% to 28\%.

Agreement that AI existential risk should be a global priority was high before the event and increased further (Wilcoxon signed-rank $p = 0.004$, LR $223{:}1$). Pre-event, 93\% of matched participants agreed to some degree (73\% strongly), rising to 96\% after (85\% strongly). The median response was ``Strongly agree'' both before and after.

We found no significant differences between participants affiliated with academia (faculty, students, and staff; $n = 25$) and those outside academia ($n = 64$) on any measure, including pre-event risk estimates, post-event shifts, confidence, and prioritization of AI risk.

\subsection{Perceived risk probability shifts stratified by prior familiarity}

\begin{figure}[p]
    \centering
    \includegraphics[width=0.48\textwidth]{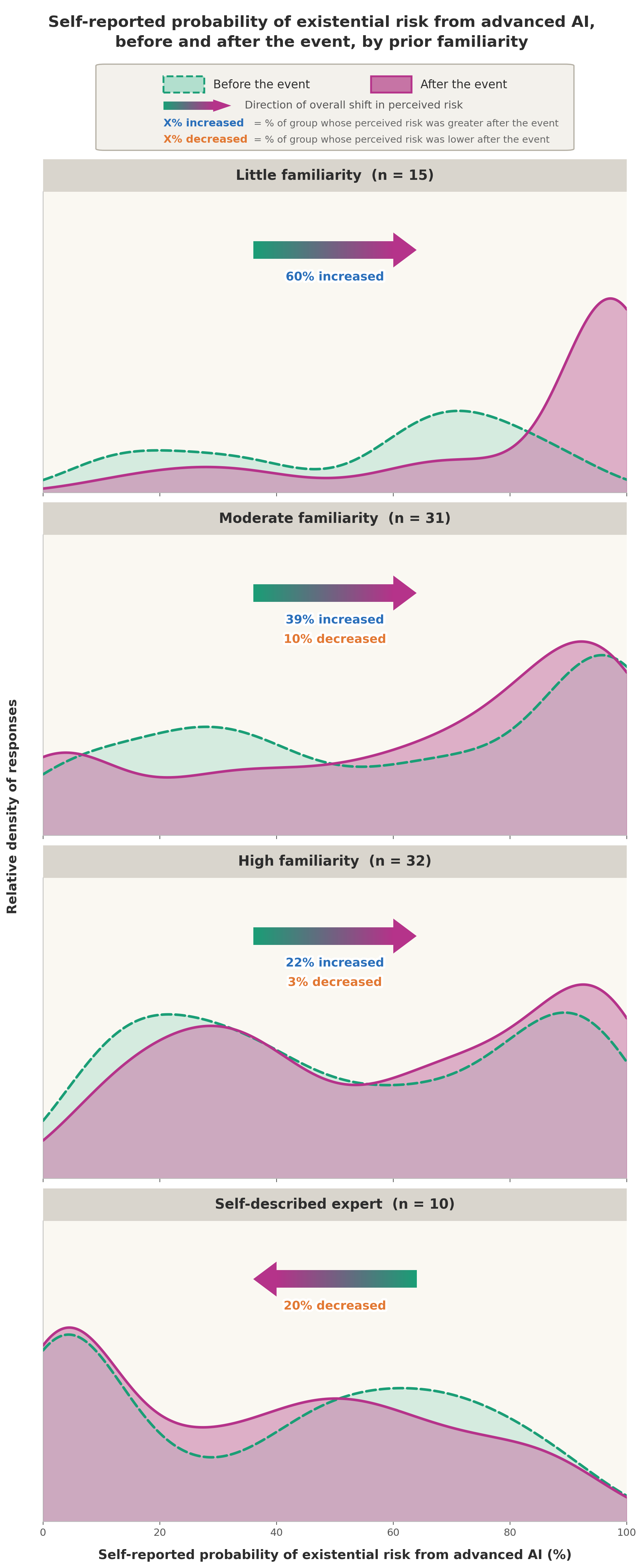}
    \caption{Changes in self-reported probability of existential risk from advanced AI developed unimpeded, before and after the event, stratified by prior familiarity. Each panel shows kernel density estimates of pre-event (teal, dashed outline) and post-event (purple/pink, solid) distributions for one familiarity group. Arrows indicate the direction of net shift, with percentages showing the fraction of participants whose estimates increased or decreased. The ``Nothing at all'' exposure group is omitted since there is only one member in that group. The gradient from ``little familiarity'' to ``expert'' reveals a monotonic decrease in upward shifts and the emergence of downward shifts.}
    \label{fig:fourpanel}
\end{figure}

The event produced a clear gradient in perceived risk probability shifts across levels of prior familiarity with AI existential risk (Figure~\ref{fig:fourpanel}, Table~\ref{tab:shifts}).

Overall, self-reported probability of AI existential risk increased significantly following the event (Wilcoxon signed-rank $W = 73$, $p < 0.001$, one-sided; likelihood ratio $3{,}734{:}1$), with a mean shift of $+6$ percentage points across all 89 matched participants. Of the 35 participants whose responses changed, 29 shifted upward and 6 shifted downward.

\begin{table}[h]
\centering
\caption{Pre/post shifts in perceived probability of AI existential risk, by prior familiarity level. Mean shift is computed from bin midpoints. Wilcoxon signed-rank tests (one-sided) and likelihood ratios (LR) are reported for each group, testing for upward shift except in the Expert row which tests for downward shift.}
\label{tab:shifts}
\begin{tabular}{lccccccc}
\toprule
Prior familiarity & $n$ & \% Up & \% Same & \% Down & Mean shift (pp) & $p$ & LR \\
\midrule
A little & 15 & 60\% & 40\% & 0\% & $+12$ & 0.002 & $512{:}1$ \\
Moderate & 31 & 39\% & 52\% & 10\% & $+10$ & $<$0.001 & $18{:}1$ \\
A great deal & 32 & 22\% & 75\% & 3\% & $+3$ & 0.023 & $13{:}1$ \\
Expert & 10 & 0\% & 80\% & 20\% & $-4$ & 0.25 & $4{:}1$ \\
\midrule
All matched & 89 & 33\% & 61\% & 7\% & $+6$ & $<$0.001 & $3{,}734{:}1$ \\
\bottomrule
\end{tabular}
\end{table}

The shift was significant within each familiarity group of sufficient size (i.e. more than ten participants). Among those with little prior familiarity ($n = 15$), 60\% reported a higher probability after the event, with no participants shifting downward ($p = 0.002$). Among those with moderate familiarity ($n = 31$), 39\% increased and 10\% decreased their probability ($p < 0.001$), with 52\% unchanged. Among those with high familiarity ($n = 32$), 22\% increased and 3\% decreased their probability ($p = 0.023$).

The expert group ($n = 10$) displayed the opposite pattern: no respondents shifted upward, and 20\% (2 of 10) shifted downward. However, with only two participants shifting, the evidence for a systematic downward trend among experts is weak ($p = 0.25$, LR $4{:}1$).

The monotonic relationship between prior exposure and probability shift was confirmed by Spearman correlation ($\rho = -0.38$, $p < 0.001$, LR $923{:}1$). Pre-event risk estimates did not significantly predict who shifted. Prior exposure level was not significantly correlated with pre-event risk estimates, indicating that all familiarity groups entered the event with similar baseline risk perceptions. After the event, however, a significant negative correlation between exposure and risk estimates emerged (Spearman $\rho = -0.25$, $p = 0.01$, LR $17{:}1$), indicating that the gradient in post-event views was created by the event rather than pre-existing.

\subsection{Confidence increased overall}

\begin{figure}[p]
    \centering
    \includegraphics[width=\textwidth]{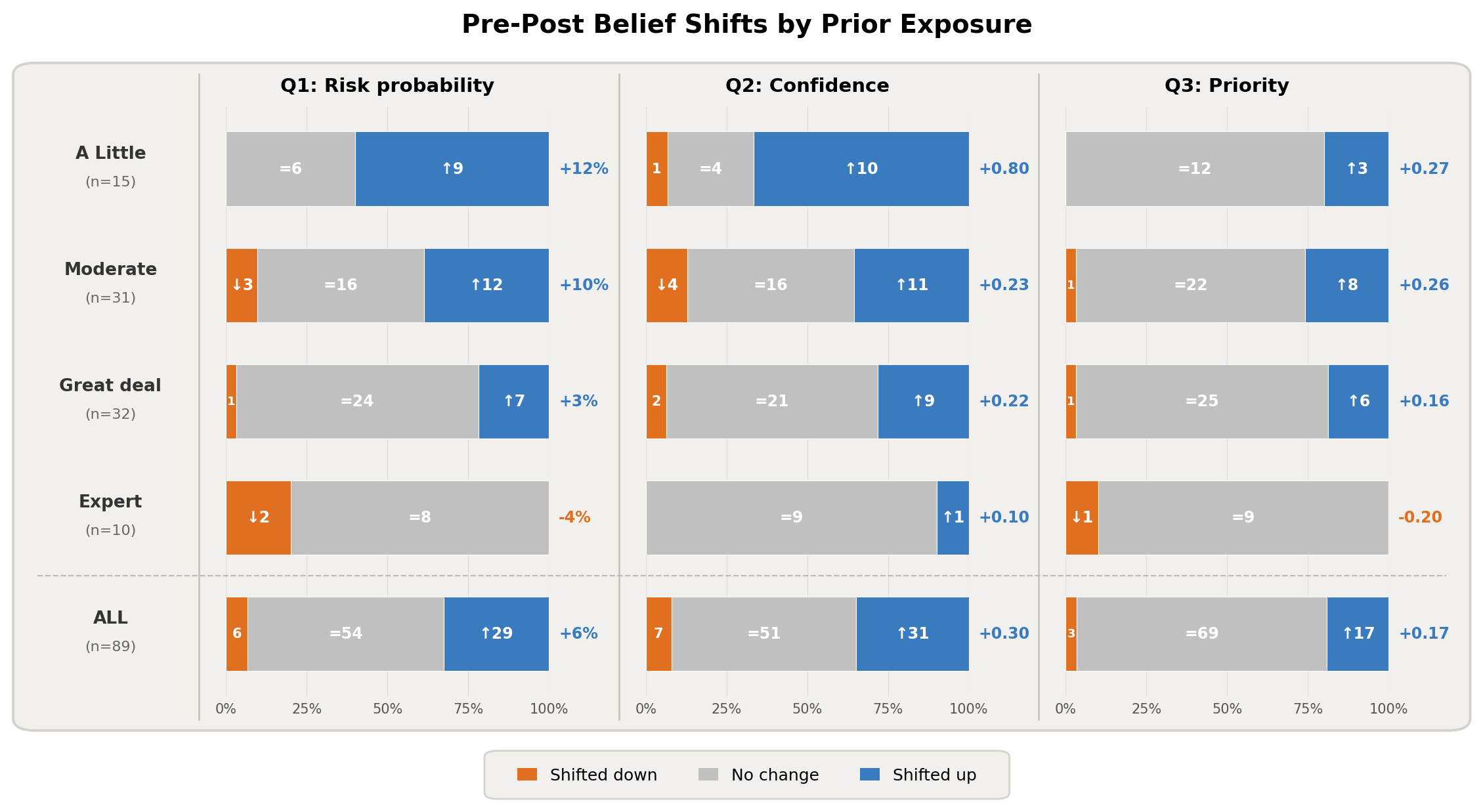}
    \caption{Direction and magnitude of individual pre--post belief shifts across all three survey questions, stratified by self-reported prior exposure to the topic of AI existential risk. Each bar shows the proportion of matched participants whose responses shifted down (orange, left), remained unchanged (gray, center), or shifted up (blue, right), with counts shown inside each segment. Right-hand labels indicate the net mean shift in percentage points (Q1) or scale points on a five-point scale (Q2, Q3). The ``Nothing at all'' exposure group is omitted since there is only one member in that group. A gradient of decreasing upward shifts with increasing prior exposure is visible across all three questions.}
    \label{fig:summary}
\end{figure}

Self-reported confidence in risk estimates increased significantly following the event (Wilcoxon signed-rank $p < 0.001$, LR $3{,}591{:}1$; Figure~\ref{fig:summary}). Across all 89 matched participants, 31 (35\%) reported higher confidence post-event, 51 (57\%) were unchanged, and 7 (8\%) reported lower confidence, yielding a mean shift of $+0.30$ on the 5-point confidence scale. Prior exposure level was negatively correlated with confidence shift ($\rho = -0.24$, $p = 0.02$, LR $14{:}1$). The largest increase was among those with little prior familiarity (mean shift $+0.80$), consistent with newcomers feeling more certain about a previously uncertain domain.

\subsection{Probability shifts correlated with confidence shifts}

Participants whose risk probability estimates shifted also tended to report increased confidence in those estimates (Spearman $\rho = 0.29$, $p = 0.006$, LR $44{:}1$). This positive correlation suggests that the observed belief changes were accompanied by a sense of greater clarity rather than increased uncertainty---consistent with belief updating.

\subsection{Comparison with the 2023 Expert Survey on AI}

The first question on our survey, about the probability of AI-caused extinction or severe disempowerment, was adapted from a 2023 survey of AI researchers \citep{grace2024}. That survey asked 2,778 AI researchers several differently framed extinction-risk questions with free-form numerical responses. The question from which ours was adapted---``What probability do you put on future AI advances causing human extinction or similarly permanent and severe disempowerment of the human species?''---received a median response of 5\% and a mean of 16.2\% ($n = 1{,}321$).

The audience at our event showed considerably higher baseline estimates: the median pre-event response was in the 40--60\% bin (midpoint 50\%) and the mean was 50.5\%, compared to a median of 5\% and mean of 16.2\% among AI researchers. This large difference likely reflects the self-selected nature of our sample, differences between the surveyed populations, the addition of an ``unimpeded'' parenthetical to our question, the use of binned versus continuous responses, and the two-plus year gap between surveys during which AI capabilities advanced considerably (see Section~\ref{sec:limitations} for further discussion).

\subsection{Comparison with national polling on global priority}

In May 2023, the Center for AI Safety (CAIS) published a one-sentence statement: ``Mitigating the risk of extinction from AI should be a global priority alongside other societal-scale risks such as pandemics and nuclear war,'' signed by hundreds of AI researchers and industry leaders including the CEOs of OpenAI, Google DeepMind, and Anthropic, as well as Turing Award laureates Geoffrey Hinton and Yoshua Bengio \citep{cais2023}.

The third question on our survey, about whether AI risk should be a global priority, uses wording identical to the CAIS statement. This same wording was used in a July 2023 YouGov poll of 1,001 U.S.\ voters conducted for the AI Policy Institute \citep{aipi2023}, and in a 2024 survey of 273 Harvard undergraduates conducted by the Harvard Undergraduate Association \citep{hirabayashi2024}. In the national poll, 70\% of respondents agreed with the statement; among Harvard undergraduates, approximately 40\% agreed. Among our matched participants ($n = 89$), 93\% agreed to some degree before the event (73\% strongly agreeing), rising to 96\% after (85\% strongly agreeing).

Direct comparison is limited by the differences in populations, timing, and context of the surveys, as well as the substantial advances in AI capabilities during the intervening period (see Section~\ref{sec:limitations}). Nonetheless, these surveys show a moderate to high degree of agreement that AI merits global attention alongside other existential risks.

\section{Discussion}

Our results demonstrate that a single public engagement event can produce measurable and statistically significant shifts in perceived AI risk ($p < 0.001$, LR $3{,}734{:}1$). The 60\% upward shift rate among those with little prior familiarity---with zero downward shifts---suggests that direct, in-person engagement with ideas about AI existential risk is most likely to shift the views of newcomers (LR $512{:}1$). The positive correlation between probability shifts and confidence shifts ($\rho = 0.29$, $p = 0.006$, LR $44{:}1$) provides additional evidence that these were belief updates: participants who changed their risk estimates also felt more confident in their revised views, rather than more uncertain.

The expert reversal is noteworthy, albeit tentative given the low sample size. Self-described experts were the only group in which the sole direction of movement was downward. One interpretation is that experts were better positioned to engage with the questions and perspectives raised during the moderated conversation, such as the question of whether large language models could carry inherent alignment advantages due to being trained on human data---a question that might resonate more with individuals who have thought deeply about AI architectures and AI alignment research.

The increase in confidence across the sample as a whole, including among those whose risk estimates did not change, suggests that the event left participants feeling more certain about their views, whether or not those views changed.

The high and increasing levels of agreement that AI risk should be a global priority (96\% of matched post-event respondents agreeing to some degree, $p = 0.004$, LR $223{:}1$) substantially exceeds agreement levels in both a national poll (70\%) and a Harvard undergraduate survey (approximately 40\%) \citep{aipi2023, hirabayashi2024}. While the gap between our sample and the national poll may partly reflect differences in the surveyed populations, the result is also consistent with the hypothesis that prioritization views have increased as AI technology has advanced. The further increase following the event suggests that direct engagement with arguments about AI risk can shift prioritization views even among audiences who already lean toward agreement.

\section{Limitations}
\label{sec:limitations}

Several limitations warrant mention, relating to the representativeness of our sample, our measurement approach, the external comparisons, and the generalizability of a single-event study.

There is likely self-selection bias in our sample given that attendees chose to attend an event with a provocative title about AI existential risk. Of the approximately 180 attendees, only a subset completed the surveys, introducing a potential nonresponse bias. Two subgroups---the expert group ($n = 10$) and the ``nothing at all'' group ($n = 1$)---are too small for robust statistical inference.

Several aspects of our measurement approach also deserve consideration. Participants who attended a talk arguing that AI poses an existential risk and then saw the same survey questions may have felt social pressure to report higher risk estimates; the anonymous polling format mitigates this somewhat but cannot eliminate it. Our use of seven response bins limits the granularity of detected changes---a one-bin shift could represent a change of 5 percentage points at the tails or 20 percentage points in the middle range. The familiarity categories are based on participants' own assessments, which may not correspond to objective knowledge levels.

The external comparisons we draw are subject to additional caveats. The 2023 Expert Survey on Progress in AI \citep{grace2024} surveyed AI researchers, whose estimates may be impacted by selection bias and domain-specific knowledge---while appreciation of technical challenges could lead to higher probability estimates, intuitions about tractability of alignment and career selection could bias toward lower estimates of potential associated dangers. Additionally, our probability question included a parenthetical asking respondents to assume AI development proceeds largely unimpeded, which may have pushed estimates upward relative to the expert survey. The 2023 YouGov poll \citep{aipi2023} surveyed a nationally representative sample of U.S.\ voters with no particular connection to the topic. The 2024 Harvard undergraduate survey \citep{hirabayashi2024} captured views of a specific demographic whose views may not generalize broadly. All three external surveys were conducted one to three years before our event, during which time AI capabilities advanced considerably with the release of models such as GPT-4, GPT-5, Claude 3, Claude 4, and subsequent generations; it is plausible that risk perceptions have shifted across all populations over this period.

Finally, this is a single event with a single speaker presenting one perspective, albeit with structured questions and a two-sided conversation. Generalizability to other formats, speakers, or audiences is unknown.

\section{Conclusion}

We present evidence that a public engagement event about AI existential risk---featuring a talk and two-sided conversation---produced statistically significant shifts in perceived risk probability ($p < 0.001$, LR $3{,}734{:}1$), confidence ($p < 0.001$, LR $3{,}591{:}1$), and global prioritization ($p = 0.004$, LR $223{:}1$). The median perceived probability of existential risk from advanced AI developed unimpeded rose from 50\% to 70\%, and 96\% of post-event respondents agreed that AI existential risk should be a global priority. The magnitude of the probability shift decreased monotonically with prior familiarity ($\rho = -0.38$, $p < 0.001$, LR $923{:}1$), and the positive correlation between probability and confidence shifts ($\rho = 0.29$, $p = 0.006$, LR $44{:}1$) suggests these changes reflected belief updating. These results contribute to a nascent literature on how public audiences process information about AI risk.

\section*{Acknowledgments}

We thank Melissa Franklin, Jeffrey Mayersohn, Marina Werbeloff, Seth Lewis, Erin Collins, Cassie Davis, Elliott Ronna, Michael Leach, and Ronald Lacey for organizing and supporting the event; the Harvard Division of Science, Harvard Book Store, and Harvard Library for co-organizing the Harvard Science Book Talk series; and the audience members who participated in the survey. We also thank Logan McCarty and Eric Moskowitz for discussions of the data. We also acknowledge the assistance of Claude (Anthropic) in the preparation of this manuscript.

\bibliographystyle{plainnat}

\end{document}